\RequirePackage{fix-cm}
\documentclass[smallextended]{svjour3}       
\smartqed  
\usepackage{appendix}
\usepackage{amsmath}
\usepackage{graphicx}
\usepackage{lineno}
\usepackage{array}
\usepackage{longtable}
\usepackage{natbib}
\usepackage{units}
\usepackage{bbold}
\usepackage{framed}
\setcitestyle{aysep={}}

\newcommand*\patchAmsMathEnvironmentForLineno[1]{%
\expandafter\let\csname old#1\expandafter\endcsname\csname #1\endcsname
\expandafter\let\csname oldend#1\expandafter\endcsname\csname end#1\endcsname
\renewenvironment{#1}%
{\linenomath\csname old#1\endcsname}%
{\csname oldend#1\endcsname\endlinenomath}}%
\newcommand*\patchBothAmsMathEnvironmentsForLineno[1]{%
\patchAmsMathEnvironmentForLineno{#1}%
\patchAmsMathEnvironmentForLineno{#1*}}%
\AtBeginDocument{%
\patchBothAmsMathEnvironmentsForLineno{equation}%
\patchBothAmsMathEnvironmentsForLineno{align}%
\patchBothAmsMathEnvironmentsForLineno{flalign}%
\patchBothAmsMathEnvironmentsForLineno{alignat}%
\patchBothAmsMathEnvironmentsForLineno{gather}%
\patchBothAmsMathEnvironmentsForLineno{multline}%
}

\begin{document}

\title{Wind profiles for WKB Prandtl models based on slope and free air flow}

 \author{Maximilian Arbeiter  \\ Iris Rammelm\"uller*  \\ Gunter Sp\"ock }

\authorrunning{M. Arbeiter, I. Rammelm\"uller and G. Sp\"ock} 

\institute{Maximilian Arbeiter \at
              University of Klagenfurt, Department of Statistics, Austria 
           \and
            Iris Rammelm\"uller \at
              *Corresponding Author\\
              University of Klagenfurt, Department of Statistics, Austria \\
              \email{Iris.Rammelmueller@aau.at} \\
              ORCID: 0000-0001-9328-9357
           \and
           Gunter Sp\"ock \at
              University of Klagenfurt, Department of Statistics, Austria
}

\date{Received: DD Month YEAR / Accepted: DD Month YEAR}

\maketitle

\begin{abstract}
In this article the WKB (Wentzel-Kramers-Brillouin) Prandtl model serves as the baseline for the study of different kinds of slope flows which can occur over inclined surfaces. 
The Prandtl-type model couples basic boundary-layer dynamics and thermodynamics for pure slope flows.
We provide an answer to the question if it is possible to obtain the matching WKB Prandtl model using only friction velocity, friction temperature, and sensible heat flux.
This instantly raises the query if there is a transition or combination between the WKB-Prandtl model for slope flows and the Monin-Obukhov similarity theory for free-air flows and vice versa.
As a result, we show the difference between friction velocity and friction temperature calculated using the Monin-Obukhov similarity theory and those computed using the WKB Prandtl model.
There is ongoing research into hill-perturbed non-neutral wind profiles because of their potential utility in numerous applications. Hence, further discussion includes how the new parametrization of the WKB Prandtl model may be used to calculate slope and free-air flows in a micro-meteorological model of an alpine valley, e.g. for pollutant dispersion calculations.

\end{abstract}

\section{Introduction}

Boundary layer meteorology, or the study of the lower part of the atmosphere, is the most significant topic of meteorology for air dispersion models. The planetary boundary layer is a part of the atmosphere whose behaviour is directly influenced by contact with the surface, e.g. \cite{Garratt1992}; \cite{DeWisher2014}. Therefore, dynamic pollutant dispersion simulation requires accurate modelling of wind speeds, wind directions, and sensible heat fluxes, e.g. \cite{Rotach2007}; \cite{DeWisher2014}. Non-neutral wind profiles play an important part in a variety of transport phenomena, such as heat and mass transfer, but also in flow-induced mixing. The flow encounters with different elevations and slopes at the surface, which lower wind speed and introduce random vertical and horizontal velocity components across the primary flow direction. The interest in hill-perturbed non-neutral flows has increased considerably in the last several decades, e.g. \cite{Egger1990}; \cite{Oerlemans1993}; \cite{Fedorovich2009}; \cite{Horvath2011}. In many practical problems, it appears that an inadequate wind profile is the most obstructive factor to a solution of the problem at hand. 
Due to changes in atmospheric stability, different kinds of slope flows can occur over inclined surfaces.  
Within the stable night-time boundary layer, katabatic down-slope flows occur, while anabatic up-hill flows occur during unstable daytime conditions. 
These slope flows are ubiquitous over complex terrain wherever underlying surface inclination exists over an appreciable distance. There are various aspects, modifications and improvements to basic and advanced models of slope flows, e.g. \cite{Mahrt1982}; \cite{Egger1990}; \cite{Smith2005}; \cite{Zardi2013}.
One of the more common analytic models for the corresponding thermally driven simple slope flows is that of Prandtl (1942). During the last 20 years, the original Prandtl model was extended by, e.g. \cite{Grisogono2001}, (\citeyear{Grisogono2000}), (\citeyear{Grisogono2002}).  \cite{Axelsen2012} discussed the strengths and limitations of the classical Prandtl model, focusing primarily on its applicability to large-eddy simulations (LES), and secondarily on its suitability for use with limited observational data sets, such as those presented by, e.g. \cite{Van1997}; \cite{Parmhed2004}; \cite{Axelsen2009}.
Field experiments have been performed for katabatic flows at the Pasterze glacier in Austria, e.g. \cite{PASTEX1994}; \cite{Grisogono2001}. Later \cite{Grisogono2015} extended the original Prandtl model by adding to the steady state Boussinesq Navier-Stokes equation a small $\epsilon$-potential temperature flux term, which enhances along-slope advection of the potential temperature $\unit[\theta]{\, [K]}$ but is counteracted by parameterized turbulent mixing, e.g. \cite{Boussinesq1897}. This is called the WKB$_{\epsilon}$ Prandtl model of order 1. \cite{Grisogono2015} showed how the friction velocity $\unit[u_{*} ]{\, [ms^{-1}]}$, friction temperature $\unit[\theta_{*}]{\, [K]}$, and sensible heat flux $\unit[Q_{H}]{\, [Jm^{-2}s^{-1}]}$ can be calculated in the WKB$_{\epsilon=0}$ Prandtl model. 
Thus, slope flows remain subject of vigorous research, e.g. \cite{Fedorovich2009}; \cite{Zardi2012}. 
\newpage
\noindent
The first objective of this article is to obtain the matching WKB$_{\epsilon}$  Prandtl model using only friction velocity, friction temperature, sensible heat flux, Prandtl number, eddy thermal diffusivity function, and slope angle.
Our second objective is to integrate various types of terrain, for which we utilize Monin-Obukhov similarity theory. This theory is then employed to develop the appropriate Prandtl model. 
The question now is whether the Monin-Obukhov similarity theory parameters $u_{*}$, $\theta_{*}$, and $Q_{H}$ derived for virtually flat terrain may also be utilised as parameters to obtain those in the Prandtl WKB$_{\epsilon}$ model, or whether they require a functional change. 
An intriguing question is whether a transition and/or combination exists that links the Monin-Obukhov similarity theory for free-air flows with the WKB Prandtl model for inclined flows, and vice versa.
Our models are validated with the WKB Prandtl model presented in \cite{Grisogono2015}, which serves as the baseline for this work. 
\newline 
The article is structured as follows: 
in Section \ref{sec:WKB} the extended Boussinesq Navier-Stokes equations and their solutions for the WKB$_{\epsilon=0}$ and WKB$_{\epsilon>0}$ Prandtl model are given. Section \ref{sec:ustern} describes how the parameters $u_{*}$, $\theta_{*}$, and $Q_{H}$ may be derived for the WKB$_{\epsilon}$ Prandtl model.
The most innovative part of this article is Sect. \ref{sec:ustern2}, where it is shown how the WKB$_{\epsilon}$ Prandtl model may be obtained from knowing only the slope angle $\unit[\alpha]{\, [rad]}$, $u_{*}$, $\theta_{*}$, $Q_{H}$, Prandtl number $Pr$, and the parametrically specified height-dependent eddy thermal diffusivity $\unit[K_{H}(z)]{\, [m^2s^{-1}]}$. Section \ref{sec:results} illustrates how in particular cases the WKB$_{\epsilon}$ Prandtl model may be accurately computed using only the slope angle, friction parameters, and sensible heat flux.
The question of the relationship among the friction parameters, $u_{*}$ and $\theta_{*}$, in flat and steep terrain will be answered in Sect. \ref{sec:translating}. In Section \ref{sec:combination} we combine the WKB$_{\epsilon}$ Prandtl model for steep terrain with the Monin-Obukhov model for flat terrain, so that the complete modelling of the mean flow affected by buoyancy over varying terrain, e.g. in an alpine valley, becomes possible.
The findings are highlighted in the concluding Sect. \ref{sec:conclusions}, and an appendix contains a symbol list.

\section{The Prandtl- and the WKB$_{\epsilon}$ Prandtl model}\label{sec:WKB}
The classic Prandtl model has been invented by \cite{Prandtl1942} and has been used by several authors, e.g. \cite{Defant1949}; \cite{Smith1979}; \cite{Egger1990}; \cite{Axelsen2009}. 
The governing steady-state one-dimensional equations in the tilted coordinate frame for a constant slope angle $\alpha$, $0<\alpha<\frac{\pi}{2}$ with quasi-hydrostatic approximation are given by, e.g. \cite{Grisogono2015}. This is a modification of the original coupled linear diffusion equations, one for momentum and one for temperature. In the presence of progressively changing background eddy coefficients, the added $\epsilon-$term in Eq. (\ref{eq:1}) turns the original system weakly nonlinear.
According to \cite{Grisogono2015}, the parameters have the usual meaning:
\begin{equation}
\begin{aligned}
  0&=g\frac{\Delta\theta}{\theta_{0}}\sin (\alpha)+K_{H}Pr\frac{\partial^{2}u}{\partial z^{2}}, \\
  0&=-\left(\Gamma_0+\epsilon\frac{\partial\Delta\theta}{{\partial z }}\right)u\sin (\alpha)+K_{H}\frac{\partial^{2}\Delta\theta}{\partial z^{2}},\label{eq:1}
\end{aligned}
\end{equation}
where $\unit[g]{\, [ms^{-2}]}$ is the acceleration due to gravity, Prandtl number $Pr=K_{M}/K_{H}$,  eddy viscosity $\unit[K_{M}]{\, [m^2s^{-1}]}$, eddy thermal diffusivity $\unit[K_{H}]{\, [m^2s^{-1}]}$, absolute temperature at surface level $\unit[\theta_{0}]{\, [K]}$, potential temperature anomaly $\unit[\Delta\theta]{\, [K]}$, and $\epsilon\in (0,1]$. The boundary conditions are selected as the usual ones, e.g. \cite{Grisogono2015}, for the Prandtl model: $\Delta\theta(z=z_{0})=C$, $u(z=z_{0})=0$, $\Delta\theta(z\rightarrow\infty)=0$, and $u(z\rightarrow\infty)=0$, where $\unit[z_0]{\, [m]}$ is the roughness length measured orthogonal to the surface and $\unit[C]{\, [K]}$ is the Prandtl amplitude. $\Delta \theta$ describes the potential temperature deficit in katabatic flows and the converse in anabatic flows with a temperature surplus, e.g. \cite{Grisogono2015}. 
\noindent 
$\unit[\Gamma_0]{\, [Km^{-1}]}$ describes the background gradient measured orthogonal to the slope and relates to the background potential temperature gradient $\unit[\Gamma]{\, [Km^{-1}]}$ as follows: 
\begin{equation*}
\Gamma_0 = \frac{\Gamma}{\cos(\alpha)}.
\end{equation*}
\noindent
Equation (\ref{eq:1}) originates from the time-independent stochastic Navier-Stokes equation, e.g. \cite{Birnir2013}, using ordinary K-closure: 
\begin{equation}
\begin{aligned}
  -\langle u,w\rangle&=K_{M}\frac{\partial u}{\partial z},\\
  -\langle\theta,w\rangle&=K_{H}\left(\frac{\partial \Delta\theta}{\partial z }+\Gamma_0\right),\label{eq:cov}
\end{aligned}
\end{equation}
whereby the potential temperature is defined as follows:
\begin{eqnarray}
\theta(z)=\theta_{0}+\Gamma_0 (z-z_0)+\Delta\theta (z),\label{eq:theta}
\end{eqnarray}
where $\unit[z]{[m]}$ varies orthogonally above the inclined surface. Note that $\langle u,w\rangle$ denotes the covariance between $u$ and $w$. 
\noindent 
The Reynolds decomposition is one of the most fundamental approaches to quantify turbulences. Therefore, the velocity can be considered as a vector with additive deterministic and random components. In other words, we can partition each of the variables into mean and turbulent parts. Thus, at the left-hand side of Eq. (\ref{eq:cov}), at the covariance, $u$, $w$, and $\theta$ have a stochastic interpretation. On the right-hand side of these equations their interpretation is deterministic as the mean wind speed gradient and mean potential temperature gradient. We only speak about wind speeds and potential temperature and assume that stochastic and deterministic interpretation can be detected from context.\\
The WKB$_{\epsilon}$ approach expands $u$ and $\Delta\theta$ into a power series of $\epsilon$ as follows, e.g. \cite{Metivier2003}: 
\begin{equation}
\begin{aligned}
  u&=u_{0}+\epsilon u_{1}+\epsilon^{2}u_{2}+...,\\
  \Delta\theta&=\Delta\theta_{0}+\epsilon\Delta\theta_{1}+\epsilon^{2}\Delta\theta_{2}+....\label{eq:3}
\end{aligned}
\end{equation}
\noindent 
$u$ is assumed to be the positive down-slope wind component in the katabatic case and assumed to be negative for the up-slope wind component in the anabatic case.
\cite{Grisogono2001} as well as \cite{Grisogono2015} selected a height-dependent eddy thermal diffusivity function for  $z\geq z_{0}$:
\begin{eqnarray}
K_{H}(z)=K_{0}\frac{z}{h}\exp\left(-\frac{z^{2}}{2h^{2}}\right), \label{eq:2}
\end{eqnarray}
which depends on two parameters related to the maximum value of $K_{H}(z)$ and the elevation $\unit[h]{[m]}$, where the maximum is attained, i.e. $(K_{0}, h)$.\\
$(u_{0},\Delta\theta_{0})$ are the solutions of the WKB$_{\epsilon=0}$ model and well known from literature to be defined as follows:
\begin{equation}
\begin{aligned}
u_{0,WKB}(z)&=-C\mu\exp (-I(z))\sin (I(z)), \\
\Delta\theta_{0,WKB}(z)&=C\exp (-I(z))\cos (I(z)),\label{eq:6}
\end{aligned}
\end{equation}
where
\begin{eqnarray*}
I(z)=\left(\frac{\sigma_{0}}{2}\right)^{\frac{1}{2}}\int\limits_{z_{0}}^{z}K_{H}(z)^{-\frac{1}{2}}dz,
\end{eqnarray*}
\noindent 
and $\sigma_{0}=N_{\alpha}Pr^{-1/2}$ with $N_{\alpha}=N\sin(\alpha)$, $N=\sqrt{\Gamma_0 g/\theta_{0}}$, and $\mu=\sqrt{g/(\theta_{0}\Gamma_0 Pr)}$. 
It's worth noting that the original Prandtl solution may be recovered by means of setting $K_{H}(z)=K_{H}$. 
\cite{Grisogono2015} showed that first order solutions in the WKB$_{\epsilon}$ model are given as follows:
\begin{equation}
\begin{aligned}
  \Delta\theta_{1,WKB}(z)&=\Delta\theta_{A,WKB}\exp(-I(z))\times\left(-\frac{1}{15}\sin (I(z))-\frac{1}{6}\cos(I(z))\right) \\
                +\Delta\theta_{A,WKB}&\exp(-2I(z))\times\left(\frac{1}{15}\sin (2I(z))+\frac{1}{15}\cos(2I(z))+\frac{1}{10}\right), \\ 
& \\  
  u_{1,WKB}(z)&=u_{A,WKB}\exp(-I(z))\times\left(-\frac{1}{3}\sin (I(z))+\frac{2}{15}\cos(I(z))\right) \\
                +u_{A,WKB}&\exp(-2I(z))\times\left(\frac{1}{30}\sin (2I(z))-\frac{1}{30}\cos(2I(z))-\frac{1}{10}\right), \label{eq:9} 
\end{aligned}
\end{equation}
where the amplitudes are
\begin{equation}
\begin{aligned}
  \Delta\theta_{A,WKB}&=\left(\frac{2}{\sigma_{0}}\right)^{\frac{1}{2}}C^{2}\mu\sin(\alpha)K_{H}(z)^{-\frac{1}{2}},\\
  u_{A,WKB}&=\left(\frac{\sigma_{0}}{2}\right)^{\frac{1}{2}}C^{2}\frac{\mu}{\Gamma_0}K_{H}(z)^{-\frac{1}{2}}.\label{eq:10}
\end{aligned}
\end{equation}
The boundary conditions for these solutions are as follows: $\Delta\theta_{1,WKB}(z_{0})=0$, $u_{1,WKB}(z_{0})=0$, $\Delta\theta_{1,WKB}(z\rightarrow\infty)=0$, and $u_{1,WKB}(z\rightarrow\infty)=0$. The total WKB$_{\epsilon}$ solutions up to the first order are given by:
\begin{equation}
\begin{aligned}
\Delta\theta_{WKB}(z)&=\Delta\theta_{0,WKB}(z)+\epsilon\Delta\theta_{1,WKB}(z),\label{eq:x0}\\
u_{WKB}(z)&=u_{0,WKB}(z)+\epsilon u_{1,WKB}(z).
\end{aligned}
\end{equation}
Based on empirical arguments, \cite{Grisogono2015} proposed to select in the katabatic case of a stable atmosphere $\epsilon=0.005$ and in the anabatic case of an unstable atmosphere $\epsilon=0.03$.
According to \cite{Metivier2003} the first order solutions (\ref{eq:x0}) are accurate  to order $\mathcal{O}(\epsilon^{2})$ to the two solutions (\ref{eq:3}) of the system of differential equations (\ref{eq:1}).


\section{Obtaining $u_{*}$, $\theta_{*}$, $Q_{H}$ in the WKB$_{\epsilon}$ Prandtl model}\label{sec:ustern}
The friction parameters $u_{*}$ and $\theta_{*}$ are defined as follows:
\begin{equation}
\begin{aligned}
  u_{*}^{2}&=\vert\langle u,w\rangle_{z=z_{0}}\vert,\label{eq:ustern2}\\
  \theta_{*}u_{*}&=-\langle \theta,w\rangle_{z=z_{0}},
\end{aligned}
\end{equation}
where the above expressions are evaluated at $z_{0}$. In the Prandtl model, $\unit[z_{j}]{\, [m]}$ specifies the height at which the maximum wind speed occurs.
One can extrapolate the momentum flux from the jet height $z_{j}$ linearly down to $z_{0}$ by means of calculating the tangent to the momentum flux at height $z_{j}$. The goal is to find the intersection of this tangent with $z_{0}$. 
Similarly, the temperature flux is extrapolated from $z_{j}$ to $z_{0}$ by assuming that it remains constant below $z_{j}$, e.g. \cite{Grisogono2001}. Thus,
\begin{equation}
\begin{aligned}
  u_{*}^{2}&=\Bigg\vert\frac{\partial(\langle u,w \rangle_{z=z_{j}})}{\partial z}\Bigg\vert(z_{j}-z_{0}),\\
  \theta_{*}u_{*}&=-\langle \theta,w\rangle_{z=z_{j}}.\label{eq:12}
\end{aligned}
\end{equation}
\noindent 
A noteworthy fact is that first and second order derivative exist in Eq. (\ref{eq:12}) since the covariances are defined as in Eq. (\ref{eq:cov}).
In the classical Prandtl model an explicit formula for the jet height $z_{j}$ is given as follows:
\begin{eqnarray*}
z_{j}=z_{0}+\pi\sqrt{\frac{K_{H}}{8\sigma_{0}}}.
\end{eqnarray*}
In the WKB$_{\epsilon\geq 0}$ model no explicit formula for $z_{j}$ can be derived and must be calculated from the wind profile. According to \cite{Grisogono2001}, explicit expressions for $u_{*}$ and $\theta_{*}$ in the WKB$_{\epsilon=0}$ and the classical Prandtl model are given as follows:
\begin{equation}
\begin{aligned}
     &u_{*}=(\vert C \vert (Pr/2)^{1/2}\mu N_{\alpha}(z_{j}-z_{0}))^{1/2}\exp(-\pi/8),&\label{eq:x1}\\
  &\theta_{*}=(\Gamma_0 K_{H}(z_{j})- \vert C \vert(N_{\alpha}Pr^{-1/2}K_{H}(z_{j}))^{1/2}\exp(-\pi/4))/u_{*}. 
\end{aligned} 
\end{equation}
We have investigated numerically whether above formulas for $u_{*}$ and $\theta_{*}$  may be used also in the WKB$_{\epsilon>0}$ case but with height $z_{j}$ calculated from the WKB$_{\epsilon>0}$ model. Hence, we have calculated derivatives by means of replacing first and second derivatives occurring in (\ref{eq:12}) by difference quotients, when using the WKB$_{\epsilon>0}$ solutions (\ref{eq:x0}). From this calculation we can see that those derivatives where $\epsilon$ occurs are almost zero. The next two examples (Table \ref{tab:1}) illustrate this fact for katabatic and anabatic cases. 
Since the deviation of $u_{*}^{2}$ from the full WKB$_{\epsilon>0}$ model to the one from the WKB$_{\epsilon=0}$ model but $z_{j}$ taken from the full model is very small, we propose to use \cite{Grisogono2001} formulas also for the WKB$_{\epsilon>0}$ model.\newline
\noindent 
The sensible heat flux is the third parameter of interest whereby \cite{Grisogono2001} recommended the following equation for $Q_H$ at height $z_{j}$ in the WKB$_{\epsilon>0}$ model as follows:
\begin{eqnarray}
 Q_{H}=-\rho c_{p}K_{H}(z_{j})\left(\frac{\partial\Delta\theta(z_{j})}{\partial z}+\Gamma_0\right),\label{eq:13}
\end{eqnarray}
\noindent 
and for the WKB$_{\epsilon=0}$ model
\begin{eqnarray}
Q_{H}&=-\rho c_{p}\theta_{*}u_{*},\label{eq:14}
\end{eqnarray}
where $\unit[c_{p}]{\, [J \ kg^{-1} \ K^{-1}]}$ is the specific heat capacity of dry air at a temperature of $\unit[20]{\, ^\circ C}$ and $\unit[\rho]{\, [kg \ m^{-3}]}$ the density of air.

\begin{table}[ht]
\begin{center}
\caption{WKB$_{\epsilon>0}$ models from \cite{Grisogono2015}, Figure 3 and 6   \label{tab:1}}  
\begin{tabular}{ll|ll}
  Par.&Units&Fig. 3&Fig. 6\\
  \hline
  $z_{0}$&$\unit[]{\, m}$&$0.15$&$0.15$\\
  $K_{0}$&$\unit[]{\, m^2s^{-1}}$&$0.49$&$9.89$\\
  $h$&$\unit[]{\,m}$&$30$&$75$\\
  $\theta_{0}$&$\unit[]{\, K}$&$273.14$&$273.14$\\
  $C$&$\unit[]{\, K}$&$-6$&$6$\\
  $\Gamma_0$&$\unit[]{\,K m^{-1}}$&$0.003$&$-0.003$\\
  $\epsilon$&$\unit[]{\,}$&$0.005$&$0.03$\\
  $\alpha$&$\unit[]{\, \text{}^{o}}$&$5^{o}$&$5^{o}$\\
  $Pr$&$\unit[]{\,}$&$2$&$2$\\
  \hline
  $u_{*}$&$\unit[]{\, ms^{-1}}$&$0.25$&$0.63$\\
  $\theta_{*}$&$\unit[]{\, K}$&$0.11$&$-0.29$\\
  $Q_{H}$&$\unit[]{\,Jm^{-2}s^{-1}}$&$-36.10$&$215.53$\\
  dev. $u_{*}^{2}$&$\unit[]{\,}$&$-1.4e^{-5}$&$-0.03$
\end{tabular}
\end{center}
\end{table}
\noindent 
In Table \ref{tab:1}, $u_{*}$ and $\theta_{*}$ are calculated with the formulas from the WKB$_{\epsilon=0}$ model using $z_j$ from the full WKB$_{\epsilon>0}$ model. The error made from non-using the full WKB$_{\epsilon>0}$ model for calculating $u_{*}^{2}$ is denoted by dev. $u_{*}^{2}$. Note that the deviation difference is always from the full model to its partial approximation. 

\section{Obtaining the WKB$_{\epsilon}$ Prandtl model from $u_{*}$, $\theta_{*}$, $Q_{H}$.}\label{sec:ustern2}
An explicit expression for the WKB${\epsilon}$ Prandtl model, relying solely on $u_{*}$, $\theta_{*}$, and $Q_{H}$, is currently unknown.
Our method uses a numerical approach to calculate the WKB$_{\epsilon}$ Prandtl model. We have written two R-functions (\cite{R}, Version $4.0.3$) to determine the model. The first R-function $F_0(K_{0},h,Q_{H})$ calculates $u_{*}$, $\theta_{*}$, and the amplitude $C$ of the Prandtl model if $K_{0}$, $h$, and $Q_{H}$ are given. The second R-function, see Fig. \ref{fig:findK0h}, calls the first one in an optimization routine whereby the Euclidean distance between the wanted $\unit[\Tilde{u}_{*}]{\, [ms^{-1}]},\unit[\Tilde{\theta}_{*}]{\, [K]}$, and the ones calculated from the first function is minimized. The minimization is done subject to the condition that the new model is proper and $Q_{H}$ is fixed at its true value.
We now describe the first function: if $K_{0}$, $h$, and $Q_{H}$ are given, Eq. (\ref{eq:13}) must be satisfied. 
To begin, substitute Eq. (\ref{eq:2}) for $K_H$ in Eq. (\ref{eq:13}) and then Eq. (\ref{eq:x0}) for the term $\Delta\theta(z_j)$ appearing in Eq. (\ref{eq:13}).
It's worth noting that $\Delta\theta_{WKB}$, from Eq. (\ref{eq:x0}), is derived by combining Eq. (\ref{eq:6}), (\ref{eq:9}), and (\ref{eq:10}).
We have a quadratic equation for $C$ after rearranging this combination of equations. This quadratic equation can be solved analytically. It has two solutions for $C$, and we take this solution as the appropriate one that is closest to the solution of the Prandtl WKB$_{\epsilon=0}$ model.
Starting with the WKB$_{\epsilon=0}$ solution for $C$ and $z_{j}$, the determination of $C$ from the above-mentioned quadratic equation is iterated. Using $z_{j}$ calculated (numerically) from (\ref{eq:x0}) with $C$ from the previous iteration, the quadratic equation (\ref{eq:13}) for $C$ is solved successively until convergence of $C$. At the end, the function calculates not only the true $C$ but also $u_{*}$, and $\theta_{*}$ from (\ref{eq:x1}). Finally, this first R-function calculates the parameters $K_{0}$, $h$, $C$, $u_{*}$, $\theta_{*}$, and $Q_{H}$.
In order to have a permissible model, \cite{Grisogono2002} argued that the WKB$_{\epsilon=0}$ models must fulfill the following condition:
\begin{eqnarray}
\max (2z_{j},z_{inv})\leq (e^{1/2}-1)h,\label{eq:cond}
\end{eqnarray}  
where $\unit[z_{inv}]{\, [m]}$ is the height measured orthogonal to the surface where the temperature inversion takes place. If the WKB$_{\epsilon=0}$ model's parameters are fixed, $z_{j}$ and $z_{inv}$ can be calculated numerically. We assume that condition (\ref{eq:cond}) must be fulfilled also for the WKB$_{\epsilon>0}$ model. As previously stated, we want to solve a quadratic optimization problem to numerically parameterize the WKB$_{\epsilon>0}$ model in the desired variables $\Tilde{u}_{*}$, $\Tilde{\theta}_{*}$. 
This can be accomplished by a relative error function applying a least squares approach in the following manner:
\begin{align}
  f(F_0(K_{0},h,Q_{H}))&= \frac{100}{\sqrt{2}}(((u_{*}-\Tilde{u}_{*})/\Tilde{u}_{*})^{2}+((\theta_{*}-\Tilde{\theta}_{*})/\Tilde{\theta}_{*})^{2}+2c^{2})^{1/2} \label{eq:f}\\
  &\longrightarrow \min\limits_{(\theta_{*},u_{*})}\nonumber
\end{align}  
with $c=0$ if condition (\ref{eq:cond}) is true and $c=0.1$ otherwise. The value $c$ ensures that we get a minimal error of $10\%$ if the condition is not fulfilled. Here, $\theta_{*}$ and $u_{*}$ are the parameters dependent on $K_{0}$, $h$, and $Q_{H}$ that can be calculated by means of our first function. Thus, we have to minimize the above function in $K_{0}$, and $h$ because according to our first function, $K_{0}$, $h$, and $Q_{H}$ uniquely determine $u_{*}$, and $\theta_{*}$. Our procedure to minimize this function is separated into two parts. It works by numerically minimizing $f$, see Eq. (\ref{eq:f}), in $K_{0}$ with $h$ fixed (the L-BFGS-B -method  is used, \cite{R}; step 2 in the function FindK0h, Fig. \ref{fig:findK0h}).
After the calculation of the optimal $K_{0}$ for several $h$-values we select the optimal tupel $(K_{0},h)$ by using an interval bisection method, e.g. \cite{Burden2015}. 
The algorithm converges and the condition (\ref{eq:cond}) is fulfilled if the function value $f$ provided by this algorithm is $<10$. Then the parameterization  of the WKB$_{\epsilon}$ Prandtl model is so that the resulting $u_{*}$, and $\theta_{*}$ deviate from the wanted $\Tilde{u}_{*}$, and $\Tilde{\theta}_{*}$ relatively by no more than $\approx 10\%$. A problem with this approach are starting values $K_{0} $ for the  L-BFGS-B--algorithm and a starting interval where the true $h$ is supposed to be found. Therefore, these parameters are application dependent, e.g. wind field in an alpine valley, katabatic wind over a glacier, and so on. We have implemented that once condition  (\ref{eq:cond}) is fulfilled for the middle $(K_{0},h)$ all corresponding subintervals use this optimal $K_{0}$-value as a starting point for the L-BFGS-B-routine.Thus, our algorithm builds-up a binary tree of suboptimal solutions
$(h, K_{0},f(F_0(K_{0},h,Q_{H})))$ among which the one with minimal $f(F_0(K_{0},h,Q_{H}))$ is selected as overall solution. This binary tree is programmed as R-function, Fig. \ref{fig:findK0h}, that recursively calls itself until termination.
\begin{figure}[ht]
\begin{framed}
\textbf{FindK0h}($h_{l},h_{r},f_{l},f_{r}$)
\begin{enumerate}
\item Set $h=\frac{h_{l}+h_{r}}{2}$.
\item $K_{0}=\arg\min\limits_{K}f(F_0(K,h,Q_{H}))$, $f_{0}=f(F_0(K_{0},h,Q_{H}))$.
\item \textbf{if} $f_{0}>\epsilon_{0}$ \textbf{then}
\begin{enumerate}  
\item \textbf{if} $f_{r}<f_{0}$ \textbf{then} $(K_{0,1},h_{1},f_{1})=$FindK0h($h,h_{r},f_{0},f_{r}$)\\
  \textbf{else} $(K_{0,1},h_{1},f_{1})=NULL$.
\item \textbf{if} $f_{l}<f_{0}$ \textbf{then} $(K_{0,2},h_{2},f_{2})=$FindK0h($h_{l},h,f_{l},f_{0}$)\\  
  \textbf{else} $(K_{0,2},h_{2},f_{2})=NULL$.
\end{enumerate}
\textbf{endif}
\item \textbf{if} $f_{r}<f_{0}$ \textbf{or} $f_{l}<f_{0}$ \textbf{then}\\
  $~~~~~~~~~~(K_{0},h,f_{0})=\min\limits_{f_{1},f_{2}}\{(K_{0,1},h_{1},f_{1}),(K_{0,2},h_{2},f_{2})\}$.
\end{enumerate}
\textbf{Return:} $(K_{0},h,f_{0})$.\\
\end{framed}
\caption{Pseudocode of the recursive algorithm for determining the $(K_{0},h)$ values of the WKB$_{\epsilon}$ Prandtl model, knowing only $\Tilde{u}_{*}$, $\Tilde{\theta}_{*}$, and $Q_{H}$ \label{fig:findK0h}}
\end{figure}
\noindent
\begin{figure}[!ht]
\centering
\includegraphics[width=1\textwidth ]{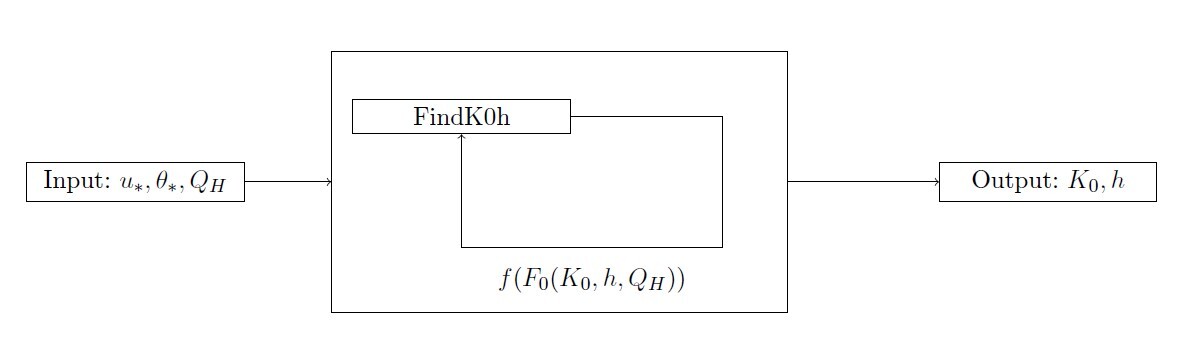}
\caption {Flowchart with the sequence of steps: The procedure takes the known parameters $u_{*}$, $\theta_{*}$, and $Q_{H}$ and forwards them to FindK0h, Fig \ref{fig:findK0h}. This recursive function dertermines $K_{0}$ and $h$ by minimizing the function $f$, Eq. (\ref{eq:f}). The auxiliary R-Function $F_{0}$ is used in every function call of FindK0h to transform the current values of $K_{0}$, $h$, and $Q_{H}$ into $u_{*}$, and $\theta_{*}$. }
\end{figure} 

\newpage
 \section{Numerical examples}\label{sec:results}
\begin{figure}[ht]
\centering
(a)\includegraphics[width=0.35\textwidth]{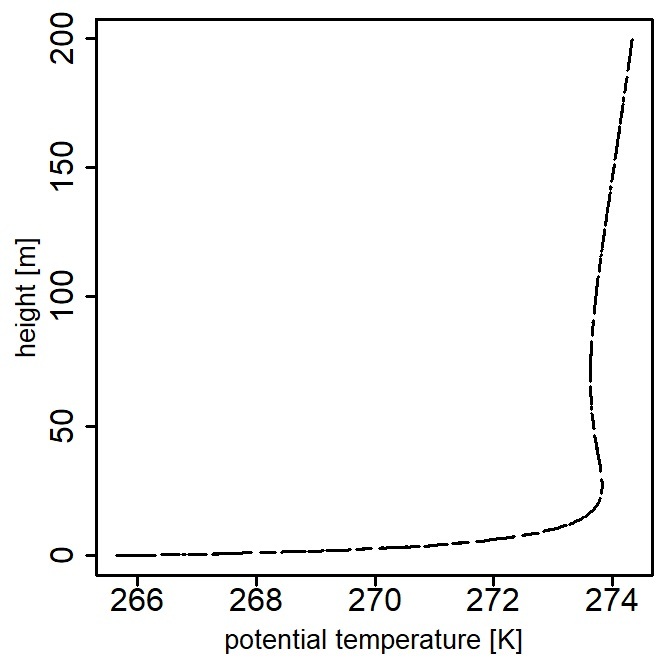}
(b)\includegraphics[width=0.35\textwidth]{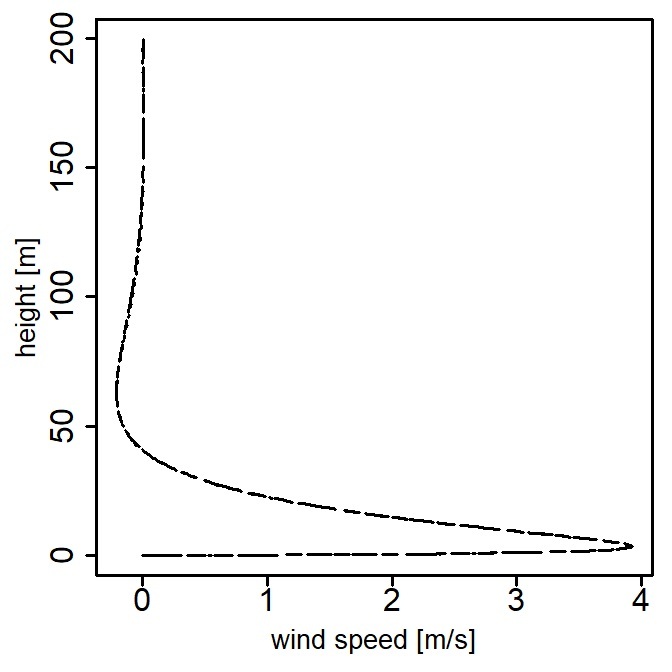}
(c)\includegraphics[width=0.35\textwidth]{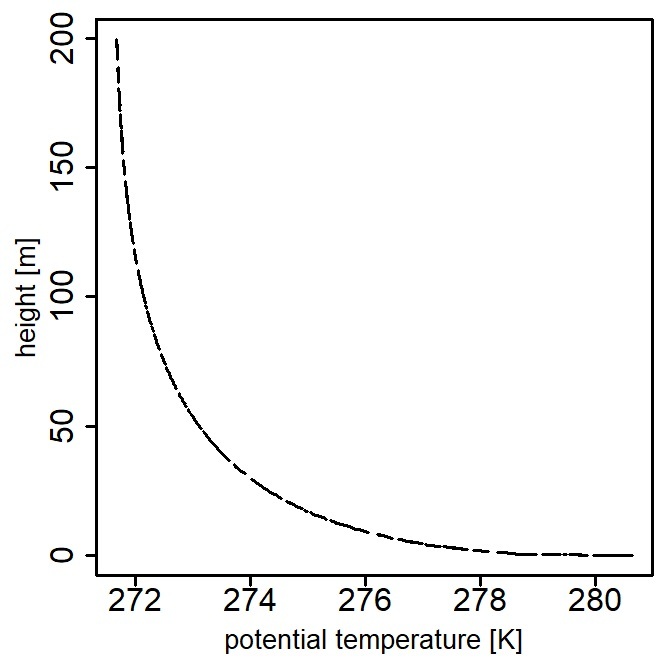}
(d)\includegraphics[width=0.35\textwidth]{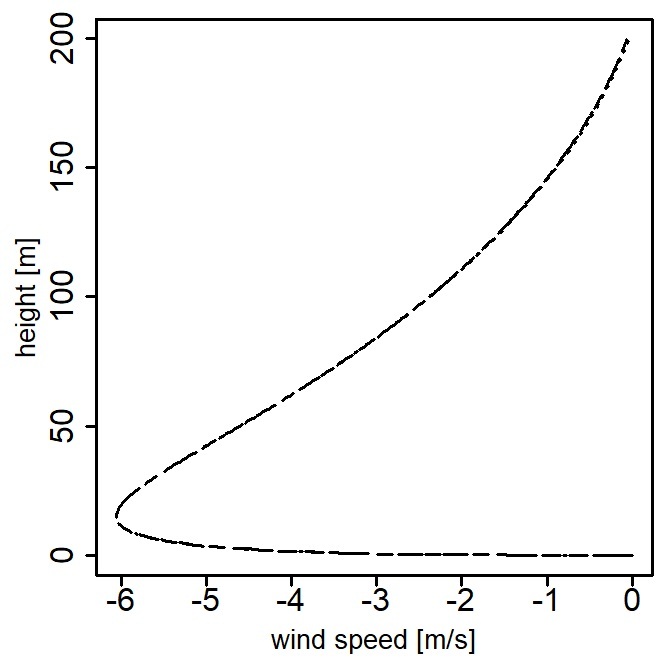}
\caption{Potential temperature and wind speed profiles for our Example 1 and 2 \label{fig:ex1}}
\end{figure}

\begin{figure}[ht]
\centering
(a)\includegraphics[width=0.35\textwidth]{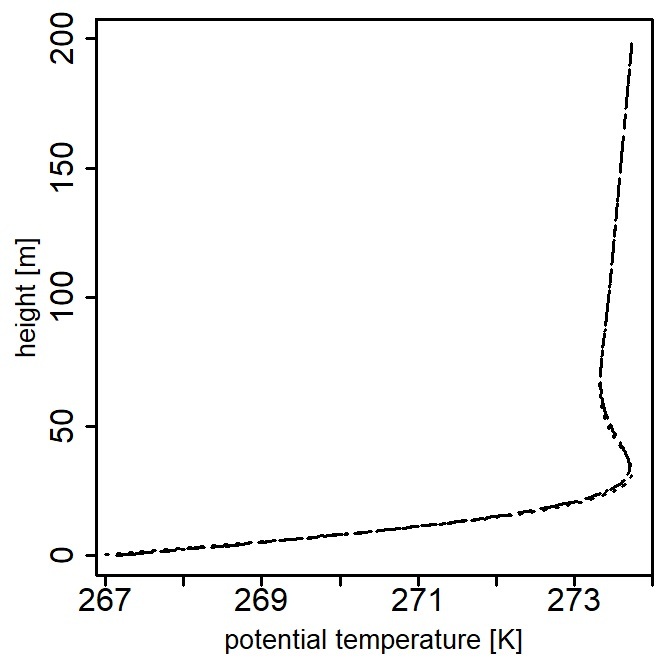}
(b)\includegraphics[width=0.35\textwidth]{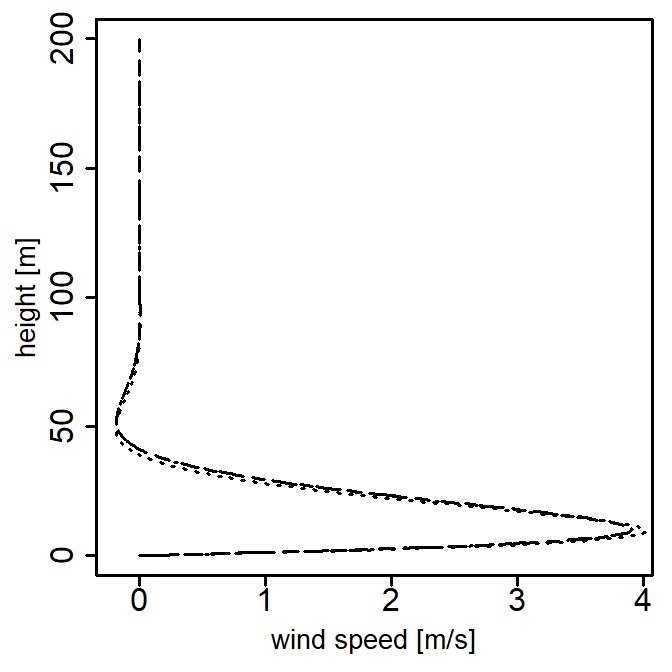}
(c)\includegraphics[width=0.35\textwidth]{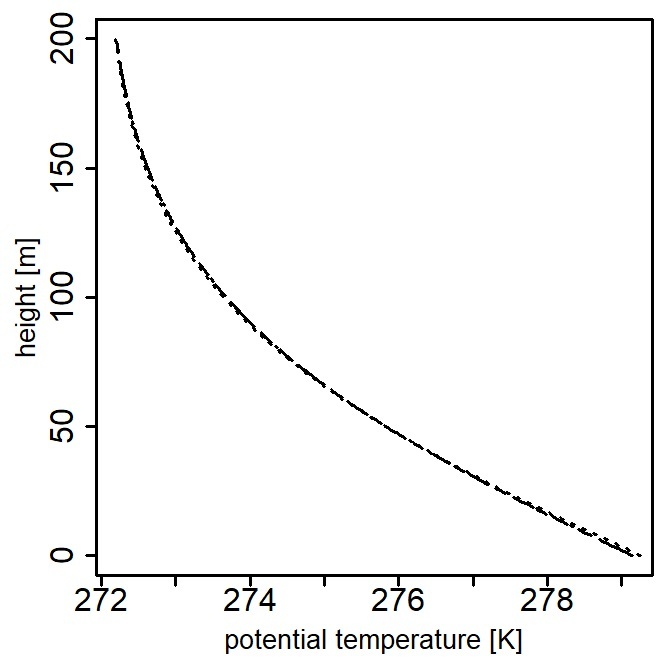}
(d)\includegraphics[width=0.35\textwidth]{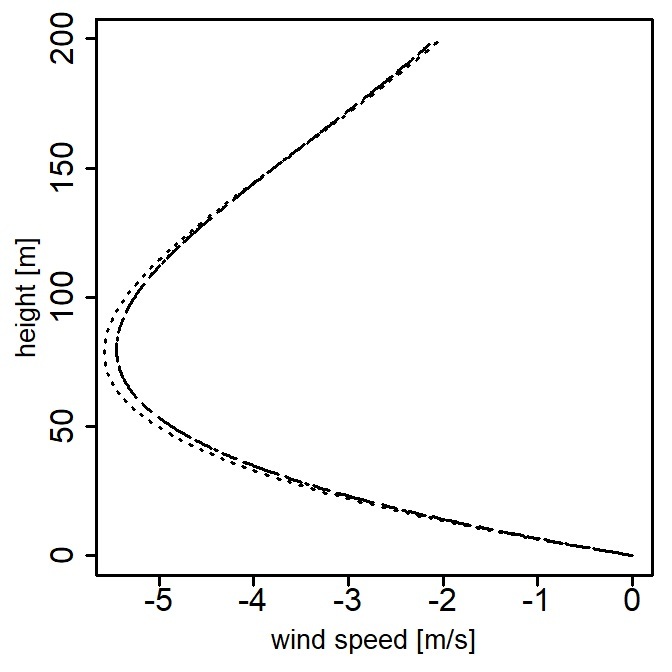}
\caption{Potential temperature and wind speed profiles whereby a, b correspond to Fig. 3 and c, d  to Fig. 6 from \cite{Grisogono2015}, WKB=FALSE \label{fig:ex3}}
\end{figure}

\begin{figure}[h!]
\centering
(a)\includegraphics[width=0.34\textwidth]{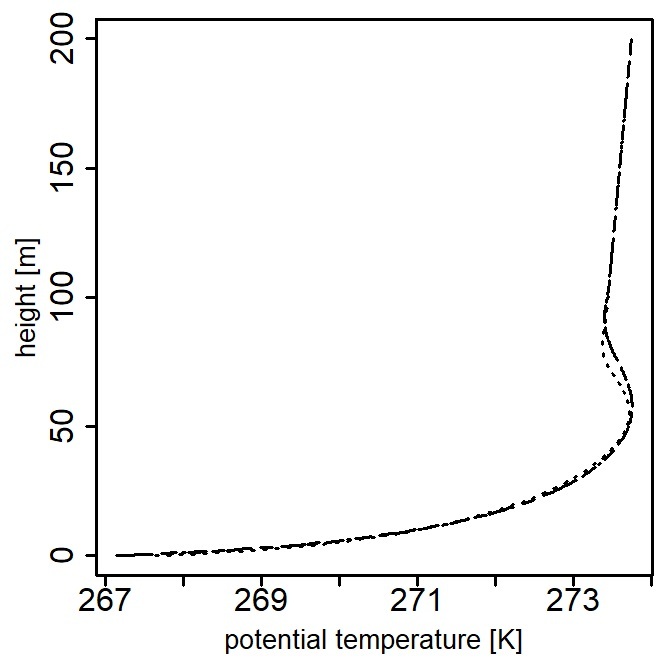}
(b)\includegraphics[width=0.34\textwidth]{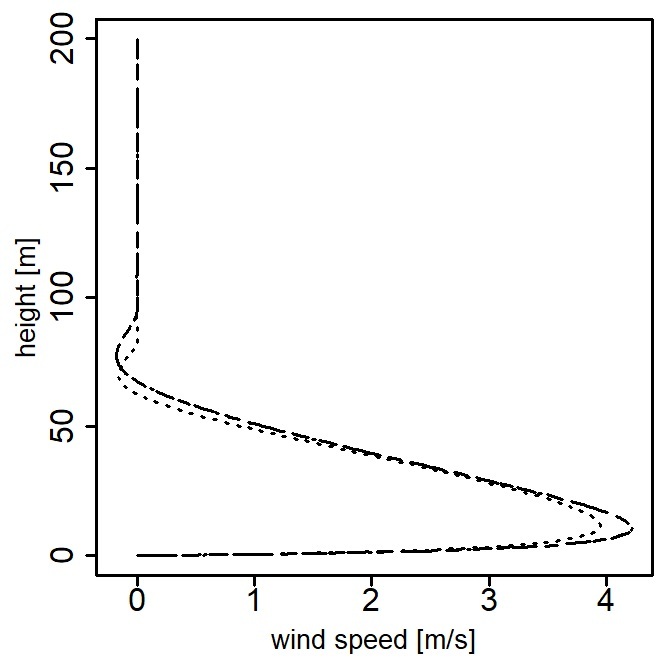}
(c)\includegraphics[width=0.34\textwidth]{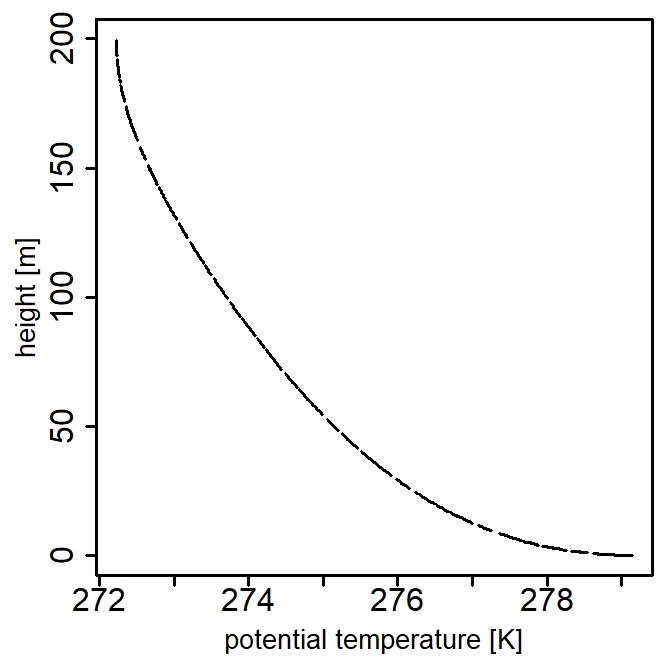}
(d)\includegraphics[width=0.34\textwidth]{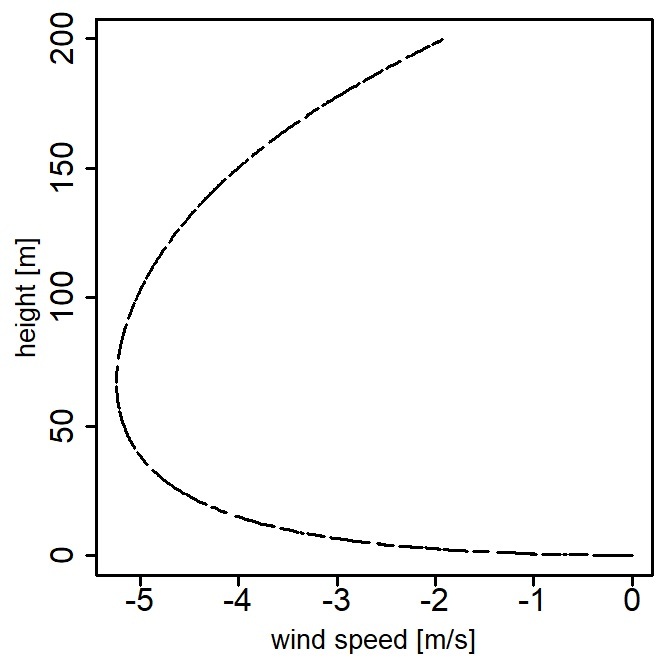}
\caption{Potential temperature and wind speed profiles whereby a, b correspond to Fig. 3 and c, d  to Fig. 6 from \cite{Grisogono2015}, WKB=TRUE \label{fig:ex4}}
\end{figure}

We provide four examples in this section that demonstrate how successfully our algorithms compute $u_{*}$, $\theta_{*}$, and $Q_{H}$. The first two examples (Table \ref{tab:2}) are our own ones and demonstrate the parameterization for the more interesting WKB=TRUE case, meaning that the diffusivity is height dependent. The next two examples (Tables \ref{tab:3} and \ref{tab:4}) are taken from Figure 3 and 6 in \cite{Grisogono2015} and demonstrate both, the WKB=FALSE and the WKB=TRUE case. \\
\noindent 
The algorithms are implemented in three R-function calls, which work as follows: 
the first call has as input $K_{0}$, $h$, and $C$ and calculates $Q_{H}$ as output and some other values, like $u_{*}$ and $\theta_{*}$. The second call uses $K_{0}$, $h$, and $Q_{H}$ (from the first call) to recalculate $C$ from the first call, and obtains $u_{*}$, and $\theta_{*}$. The last call uses $u_{*}$, $\theta_{*}$, and $Q_{H}$ (all from the first call) to recalculate $K_{0}$, $h$, and $C$. 
In below tables inputs are marked with (i) and outputs with (o). Outputs which may differ between the three calls are written down multiple times. Among the already mentioned calculated outputs, $u_{*}$, $\theta_{*}$, $Q_{H}$, $C$, $K_{0}$, and $h$ the other outputs are $u(z_{j})$, $z_{j}$, $f(K_{0},h,Q_{H})$, $z_{inv}$, and if condition (\ref{eq:cond}) is satisfied. Note that condition (\ref{eq:cond}) is always satisfied in Examples 1 and 2, but not satisfied in the WKB=TRUE case in the Grisogono examples. This is also visible from the optimum function values $f$, which then are larger than 10. In all tables it is visible from the function values $f$ and the $u_{*}$, $\theta_{*}$, and $Q_{H}$ values that our reparametrization approach works. This is visible from Figures \ref{fig:ex1}, \ref{fig:ex3}, and \ref{fig:ex4} which show wind speeds and potential temperatures before and after the reparametrization corresponding to Table \ref{tab:2}, \ref{tab:3}, and \ref{tab:4}.
Thus, in the mentioned figures only those corresponding to columns 1 and 2 in the tables give fully consistent wind speeds and potential temperatures. According to Sect. \ref{sec:ustern} wind speeds and potential temperatures corresponding to column 3 are not fully consistent, but sufficient for our purposes.\\
The convergence of the third R-function has some slight dependence on the starting value $K_{0}$ and on the proposed interval for the $h$-value which is used in the L-BFGS-B optimization routine. 
The $Q_{H}$-value calculated from the first function remains constant throughout all two other function calls.\\
\noindent 
Figure \ref{fig:ex1}a, c illustrate the potential temperature profile and Fig. \ref{fig:ex1}b, d the wind speed profile for the first and second example. The corresponding values are given in Table \ref{tab:2}. Figure \ref{fig:ex1} demonstrates that the outputs are almost identical for all three function calls in both examples. Furthermore, the relative error of $u_{*}$ and $\theta_{*}$, denoted by function value $f$, is very small and the good convergence of the algorithm can also be seen in this table. 
\begin{table}[ht]
\begin{center}
\caption{Input and output parameters for our Example 1 and 2 using WKB=TRUE \label{tab:2}}  
\begin{tabular}{ll|ll}
  Par.&Units&Ex. 1& Ex. 2\\
  \hline
  $z_{0}$(i/i/i)&$\unit[]{\, m}$&$0.0044$&$0.0044$\\
  $\theta_{0}$(i/i/i)&$\unit[]{\, K}$&$273.14$&$273.14$\\
  $\Gamma_0$(i/i/i)&$\unit[]{\, Km^{-1}}$&$0.006$&$-0.006$\\
  $\epsilon$(i/i/i)&$\unit[]{\, }$&$0.005$&$0.03$\\
  $\alpha$(i/i/i)&$\unit[]{\, \text{}^{o}}$&$5.72$&$5.72$\\
  $Pr$(i/i/i)&$\unit[]{\, }$&$1.4$&$1.4$\\
 $h $&$\unit[]{\, m}$&$\in\left[ z_0,200 \right]$&$\in\left[ z_0,200 \right]$\\
  \hline
  $u_{*}$(o/o/o)&$\unit[]{\, ms^{-1}}$&$0.17/0.17/0.17$&$0.36/0.36/0.36$\\
  $\theta_{*}$(o/o/o)&$\unit[]{\, K}$&$0.13/0.13/0.13$&$-0.35/-0.35/-0.35$\\
  $Q_{H}$(o/i/i)&$\unit[]{\, Jm^{-1}s^{-1}}$&$-29.65$&$139.95$\\
  $u(z_{j})$(o/o/o)&$\unit[]{\, ms^{-1}}$&$3.92/3.92/3.92$&$-6.05/-6.05/-6.05$\\
  $z_{j}$(o/o/o)&$\unit[]{\, m}$&$3.50/3.50/3.50$&$15.004/15.004/15.004$\\
  $C$(i/o/o)&$\unit[]{\, K}$&$-7.50/-7.49/-7.49$&$7.50/7.50/7.50$\\
  $K_{0}$(i/i/o)&$\unit[]{\, m^2s^{-1}}$&$1.25/1.25/1.04$&$8.25/8.25/8.05$\\
  $h$(i/i/o)&$\unit[]{\, m}$&$120/120/100.0022$&$120/120/117.18$\\
  $f$(-/-/o)&$\unit[]{\, \% }$&$0.0005$&$0.0099$\\
  $z_{inv}$(o/o/o)&$\unit[]{\, m}$&$27.0044/27.0044/27.0044$&$>200/>200/>200$\\
  Cond.\ref{eq:cond}&$\unit[]{\, }$&TRUE/TRUE/TRUE&TRUE/TRUE/TRUE 
\end{tabular}
\end{center}
\end{table}
\noindent 
\newpage
\noindent 
\begin{table}[ht]
\begin{center}
\caption{Input and output parameters for Fig. 3 and 6 in \cite{Grisogono2015} using WKB=FALSE \label{tab:3}}  
\begin{tabular}{ll|ll}
  Par.&Units&Fig. 3& Fig. 6\\
  \hline
  $z_{0}$(i/i/i)&$\unit[]{\, m}$&$0.15$&$0.15$\\
  $\theta_{0}$(i/i/i)&$\unit[]{\, K}$&$273.14$&$273.14$\\
  $\Gamma_0$(i/i/i)&$\unit[]{\, Km^{-1}}$&$0.003$&$-0.003$\\
  $\epsilon$(i/i/i)&$\unit[]{\, }$&$0.005$&$0.03$\\
  $\alpha$(i/i/i)&$\unit[]{\, \text{}^{o}}$&$5$&$5$\\
  $Pr$(i/i/i)&$\unit[]{\, }$&$2$&$2$\\
  \hline
  $u_{*}$(o/o/o)&$\unit[]{\, ms^{-1}}$&$0.24/0.24/0.24$&$0.69/0.69/0.69$\\
  $\theta_{*}$(o/o/o)&$\unit[]{\, K}$&$0.069/0.069/0.068$&$-0.18/-0.18/-0.18$\\
  $Q_{H}$(o/i/i)&$\unit[]{\, Jm^{-2}s^{-1}}$&$-22.06$&$145.26$\\
  $u(z_{j})$(o/o/o)&$\unit[]{\, ms^{-1}}$&$3.90/3.90/4.01$&$-5.45/-5.45/-5.58$\\
  $z_{j}$(o/o/o)&$\unit[]{\, m}$&$10.15/10.15/9.65$&$80.15/80.15/78.65$\\
  $C$(i/o/o)&$\unit[]{\, K}$&$-6/-6/-6.30$&$6/6/6.11$\\
  $K_{0}$(i/i/o)&$\unit[]{\, m^2s^{-1}}$&$0.06/0.06/0.05$&$3/3/2.89$\\
  $f$(-/-/o)&$\unit[]{\, \%}$&$0.10$&$0.04$
\end{tabular}
\end{center}
\end{table}
\noindent 
\newline
\newpage
\noindent 
Figure \ref{fig:ex3}a, b illustrate the potential temperature and wind speed profile for Fig. 3 in \cite{Grisogono2015} and Fig. \ref{fig:ex3}c, d illustrate the potential temperature and wind speed profile for Fig. 6 in \cite{Grisogono2015}. Figure \ref{fig:ex3} shows that there is a slight difference in the outputs between the function calls. This difference is visible from the terminating function value $f$ in Table \ref{tab:3}. \\
\noindent 
Figure \ref{fig:ex4}a, b illustrate the potential temperature and wind speed profile for Fig. 3 in \cite{Grisogono2015} and Fig. \ref{fig:ex4}c, d illustrate the potential temperature and wind speed profile for Fig. 6 in \cite{Grisogono2015}. Figure \ref{fig:ex4} shows that there is a difference in the outputs between the function calls. Because condition (\ref{eq:cond}) is not fulfilled, the terminating function value $f$ is around $10\%$, as shown in Table \ref{tab:4}.
\noindent 
\newpage
\begin{table}[ht]
\begin{center}
\caption{Input and output parameters for Fig. 3 and Fig. 6 in \cite{Grisogono2015} using WKB=TRUE\label{tab:4}}
\begin{tabular}{ll|ll}
  Par.&Units&Fig. 3& Fig. 6\\
  \hline
  $z_{0}$(i/i/i)&$\unit[]{\, m}$&$0.15$&$0.15$\\
  $\theta_{0}$(i/i/i)&$\unit[]{\, K}$&$273.14$&$273.14$\\
  $\Gamma_0$(i/i/i)&$\unit[]{\, Km^{-1}}$&$0.003$&$-0.003$\\
  $\epsilon$(i/i/i)&$\unit[]{\, }$&$0.005$&$0.03$\\
  $\alpha$(i/i/i)&$\unit[]{\, \text{}^{o}}$&$5$&$5$\\
  $Pr$(i/i/i)&$\unit[]{\, }$&$2$&$2$\\
$h  $&$\unit[]{\, m}$&$\in\left[ z_0,200 \right]$&$\in\left[ z_0,200 \right]$\\
  \hline
  $u_{*}$(o/o/o)&$\unit[]{\, ms^{-1}}$&$0.25/0.25/0.25$&$0.63/0.63/0.63$\\
  $\theta_{*}$(o/o/o)&$\unit[]{\, K}$&$0.11/0.11/0.11$&$-0.29/-0.29/-0.29$\\
  $Q_{H}$(o/i/i)&$\unit[]{\, Jm^{-2}s^{-1}}$&$-36.10$&$215.53$\\
  $u(z_{j})$(o/o/o)&$\unit[]{\, ms^{-1}}$&$4.21/4.21/3.947$&$-5.24/-5.24/-5.24$\\
  $z_{j}$(o/o/o)&$\unit[]{\, m}$&$10.65/10.65/11.65$&$67.15/67.15/67.15$\\
  $C$(i/o/o)&$\unit[]{\, K}$&$-6/-6/-5.50$&$6/6/6$\\
  $K_{0}$(i/i/o)&$\unit[]{\, m^2s^{-1}}$&$0.49/0.49/0.47$&$9.89/9.89/9.89$\\
  $h$(i/i/o)&$\unit[]{\, m}$&$30/30/25.075$&$75/75/75.075$\\
  $f$(-/-/o)&$\unit[]{\, \% }$&$10.0076$&$10.00001$\\
  $z_{inv}$(o/o/o)&$\unit[]{\, m}$&$57.65/57.65/55.65$&$197.15/197.15/197.15$\\
  Cond.\ref{eq:cond}&$\unit[]{\, }$&FALSE/FALSE/FALSE&FALSE/FALSE/FALSE
\end{tabular}
\end{center}
\end{table}

\newpage

\section{Translating $u_{*}$ , $\theta_{*}$, and $Q_H$ for flat terrain to those for slopes and the WKB$_{\epsilon}$ Prandtl model}\label{sec:translating}
To calculate slope winds in an alpine valley for a pollutant dispersion model in micro-meteorology, the WKB$_{\epsilon}$ Prandtl model is required.
What we get from our model in the first step are $u_{*}, \theta_{*}, Q_{H}$ from Monin-Obukhov similarity theory for flat terrain. In this part, we illustrate how to convert these values into those for the WKB$_{\epsilon}$ Prandtl model in steep terrain.\\
We make the following conventions: $u_{*}, \theta_{*}$, and $Q_H$ are the variables for flat terrain and $u_{0_{*}}, \theta_{0_{*}}$, and $Q_{H,0}$ for steep terrain with slope angle $0<\alpha<\pi/2$. \\
\noindent 
Furthermore, we assume that $u_{*}, \theta_{*}, Q_{H}, u_{0_{*}}, \theta_{0_{*}}$, and $Q_{H,0}$ are defined as usual:
\begin{equation}
\begin{aligned}
  u_{*}^{2}&=\vert\langle u,w\rangle_{z_{1}}\vert,\\
  \theta_{*}u_{*}&=-\langle \theta,w\rangle_{z_{1}},\\
  Q_{H}&=-\rho c_{p}\theta_{*}u_{*},\\
  u_{0_{*}}^{2}&=\vert\langle u_{0},w_{0}\rangle_{z_{0}}\vert,\\
  \theta_{0_{*}}u_{0_{*}}&=-\langle \theta_{0},w_{0}\rangle_{z_{0}},\\
  Q_{H,0}&=-\rho c_{p}\theta_{0_{*}}u_{0_{*}}, \label{eq:Q,u,theta}
\end{aligned}
\end{equation}
where $z_{0}=z_{1} \cdot \cos(\alpha)$ is the roughness length of the  WKB$_{\epsilon}$ Prandtl model and $z_{1}$ is the roughness length for the flat terrain. \\
\begin{figure}[ht]
\centering
\includegraphics[width=0.5\textwidth]{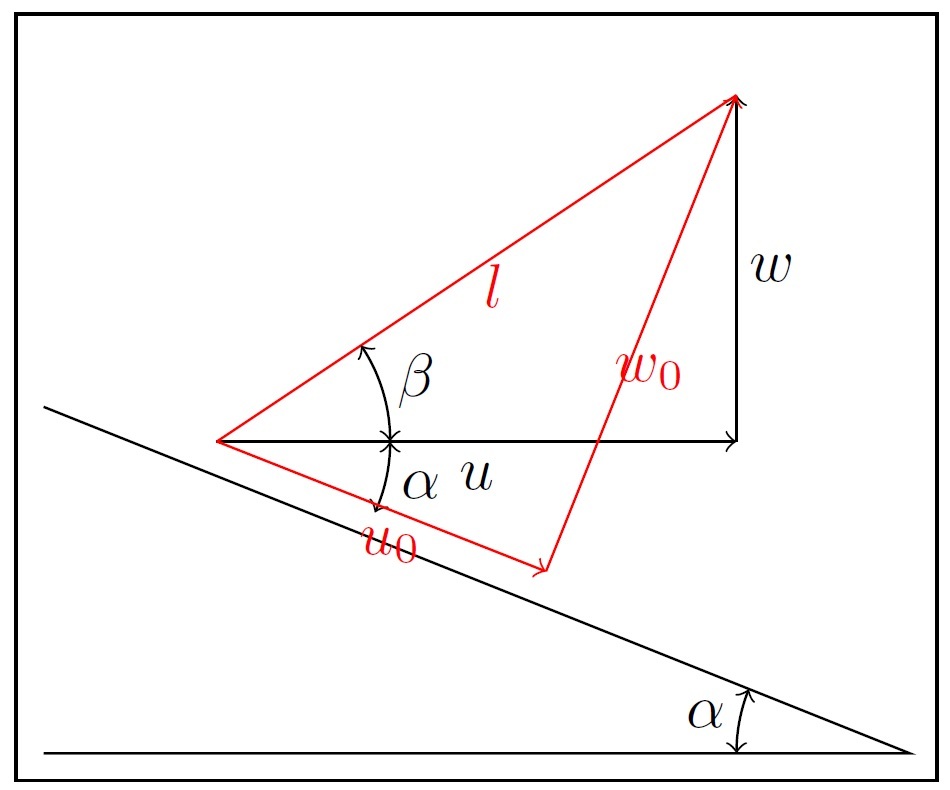}
\caption{Relation between residual wind $u,w$ and $u_{0},w_{0}$ in a katabatic case \label{fig:angles}}
\end{figure}
\noindent 
\newline 
\noindent 
Vertical up-winds are negative, while vertical down-winds are positive. Also, winds down the hill parallel to the slope are positive, winds up the hill parallel to the slope are negative. Winds up orthogonal to the slope are negative, those down are positive. 
We can deduce the following relationship since the sensible heat flux is always normal to the surface: 
\begin{equation}
Q_{H,0} = \cos(\alpha) Q_H. \label{eq:Q_H_0}
\end{equation}
\noindent 
Considering (\ref{eq:Q,u,theta}) and (\ref{eq:Q_H_0}) leads to the following relation:
\begin{equation}
\begin{aligned}
\theta_{0_{*}}u_{0_{*}}&= \cos(\alpha) \theta_{*}u_{*}. \label{eq:u_theta_ster_reation}
\end{aligned}
\end{equation}
\noindent 
In order to find relationships between $u_{*}$, $\theta_{*}$ and $u_{0_{*}}$, $\theta_{0_{*}}$ we assume a normally distributed wind vector residual $(u,w)$ at $z_1$ with covariance matrix $\Sigma$. We must preserve this covariance matrix but transfer it into the new coordinate system if we wish to derive the covariance matrix of $(u_0,w_0)$ at $z_0$. This can be accomplished in four steps: \\
First: Define the covariance matrix of $(u,w)$ at $z_1$ as follows:
\begin{equation*}
\Sigma =
\begin{pmatrix}
\sigma_{u}^2  & &   u_{*}^{2} \\
 &\\
\ u_{*}^{2}  & &  \sigma_w^2  \\
\end{pmatrix},
\end{equation*}
\noindent 
where $\sigma_u^2$ and $\sigma_w^2$ are the variances of $(u,w)$ at $z_1$. Note that the parameters $\sigma_u^2$ and $\sigma_w^2$ depend on the Pasquill-Gifford stability classes, and mixing layer height $\unit[h_{mix}]{ [m]}$ and can be calculated using the standard formulas given, e.g. \cite{DeWisher2014}. \\
\noindent
Second: Calculate the eigendecomposition of $\Sigma$ resulting in:
\begin{equation*}
\Sigma = V \Lambda V^\top ,
\end{equation*}
where $V$ is the matrix of orthonormal eigenvectors and $\Lambda$ is the diagonal matrix of the corresponding eigenvalues. \\
Third: The covariance matrix in the new coordinate system is obtained by rotating the system by $\alpha$ degrees: 
\begin{equation*}
V_0=
\begin{pmatrix}
\cos(\alpha)  & &  -\sin(\alpha) \\
 &\\
\sin(\alpha)  & &  \cos(\alpha)  \\
\end{pmatrix} \ V.
\end{equation*}
\noindent 
From this we get the covariance matrix in the new coordinate system as follows:
\begin{equation*}
\Sigma_0 = V_0 \Lambda V_0^\top =
\begin{pmatrix}
\sigma_{u_0}^2  & &   u_{0_{*}}^{2} \\
 &\\
\ u_{0_{*}}^{2}  & &  \sigma_{w_0}^2  \\
\end{pmatrix}.
\end{equation*}
\noindent 
This covariance matrix shows the change from $u_{*}^{2}$ to $u_{0_{*}}^{2}$ in the Prandtl system. \\
\noindent
Fourth: We are now ready to calculate $\theta_{0_{*}}$ from (\ref{eq:u_theta_ster_reation}), resulting in: 
\begin{equation*}
\theta_{0_{*}}=  \cos(\alpha) \theta_{*}\frac{u_{*} }{u_{0_{*}}}.
\end{equation*}
\noindent

\section{Combining models for steep and flat terrain}\label{sec:combination}
\begin{figure}[ht]
\centering
(a)\includegraphics[width=0.45\textwidth]{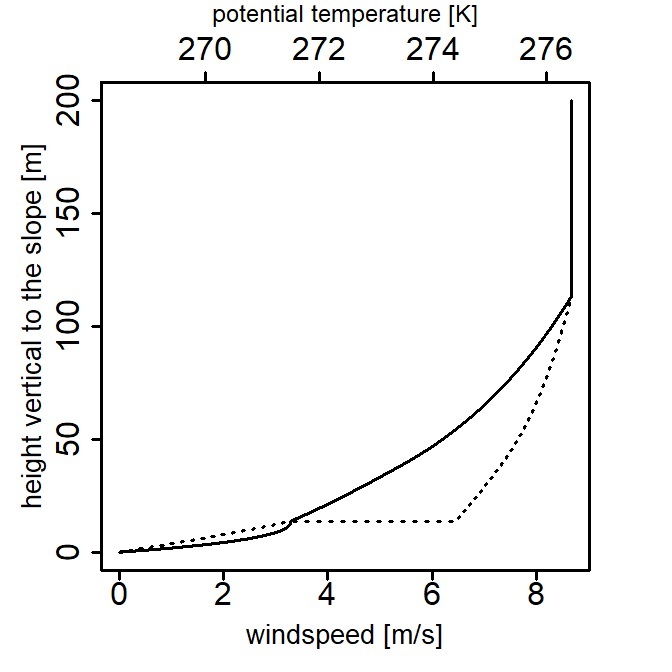}
(b)\includegraphics[width=0.45\textwidth]{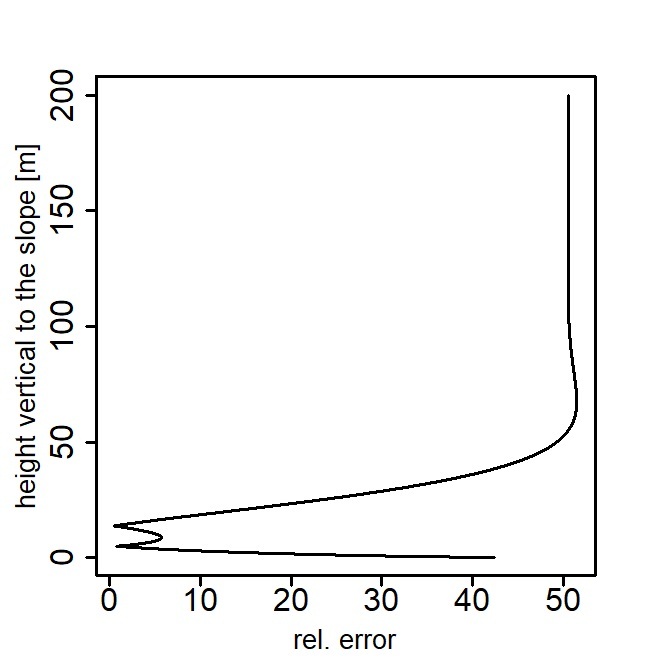}
\caption{(a): WKB$_{\epsilon}$ Prandtl model (bold line: wind speed, dotted line: potential temperature) from Fig. \ref{fig:ex3}a, b, consistently combined with the standard model for flat terrain from Monin-Obukhov similarity theory;  (b): Relative error $\unit[ ]{\, [\%]}$ calculated with (\ref{eq:f_opt}) \label{fig:Relerror}}
\end{figure}

\begin{figure}[h!]
\centering
(a)\includegraphics[width=1\textwidth]{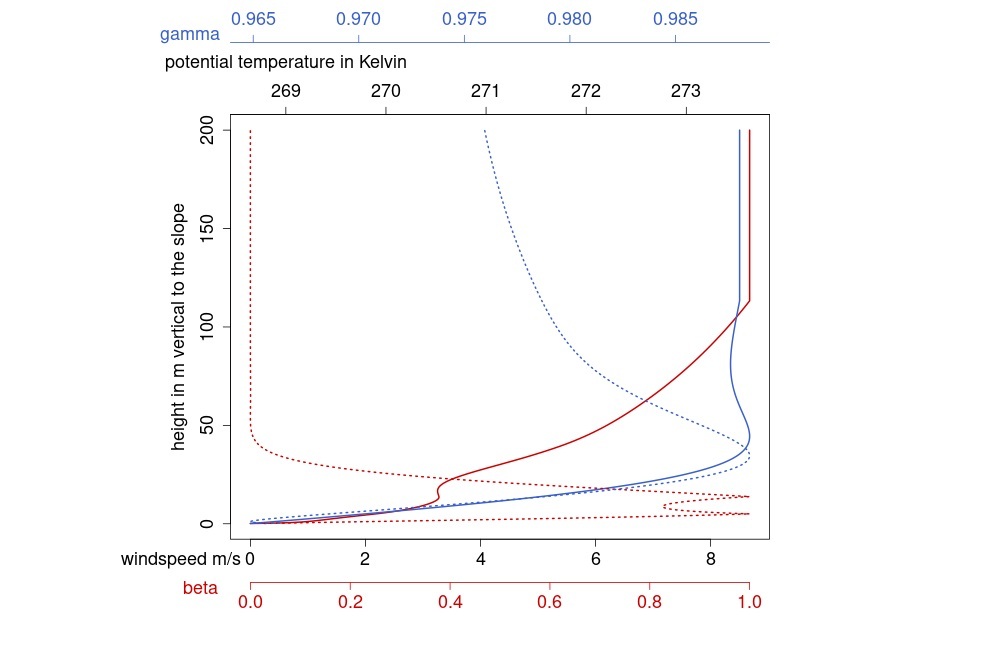}
(b)\includegraphics[width=1\textwidth]{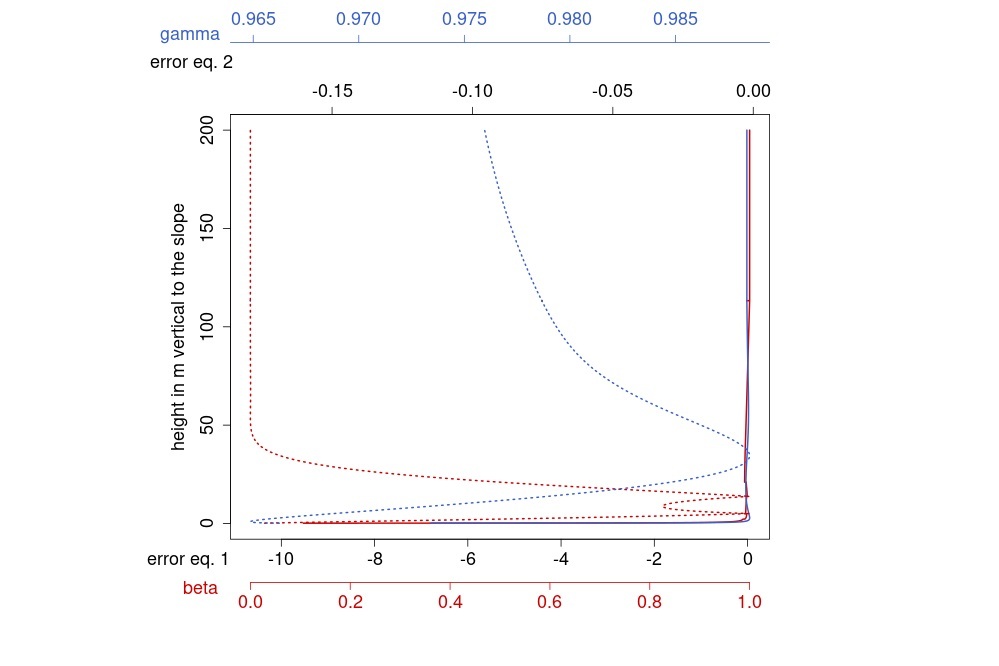}
\caption{(a): The red bold line depicts the wind speed, blue bold line the potential temperature, red dotted line depicts the weight $\beta$ and the blue dotted line depicts the weight $\gamma$; (b): The red bold line depicts the error for the first Eq. (\ref{eq:combined_ns}), blue bold line the error for the second Eq. (\ref{eq:combined_ns}), red dotted line depicts the weight $\beta$ and the blue dotted line depicts the weight $\gamma$ \label{fig:stoch_navier}}
\end{figure}

Our own application of the theory discussed earlier involves using it to calculating pollutant dispersion in an alpine valley.
The key difficulty now is how to merge the models for flat and steep terrain. We must first calculate the height at which these two models overlap. We call this height Prandtl-layer height and calculate it as follows:
\begin{equation}
 h_{0}=\arg\min\limits_{z_{*}}\left\{\frac{100}{2}\Biggl( \frac{ |u_{WKB}(z) - u_{MO}(z_{*})|}{|u_{MO}(z_{*})|}+\frac{|\theta_{WKB}(z) - \theta_{MO}(z_{*})|}{|\theta_{MO}(z_{*})|}\Biggl)\right\},\label{eq:h0}
\end{equation}
where $z_{*}=z/\cos{(\alpha)}$; $u_{WKB}(z)$ is taken from (\ref{eq:x0}) and $\theta_{WKB}(z)$ from (\ref{eq:theta}) and (\ref{eq:x0}). $u_{MO}(.)$ and $\theta_{MO}(.)$ are corresponding parameters for flat terrain based on Monin-Obukhov similarity theory, e.g. \cite{DeWisher2014}. Eq. (\ref{eq:h0}) minimizes the mean relative error between $u_{WKB}(z), u_{MO}(z_{*})$, and $\theta_{WKB}(z), \theta_{MO}(z_{*})$ to get $h_0$. \\
\noindent 
We now are able to define a combined model: For $z_{*}\leq h_{0}$ we assume the WKB$_{\epsilon}$ Prandtl model to be true; for $z_{*}>h_{0}$ we accept the Monin-Obukhov similarity theory for flat terrain. An illustrating example can be seen in Fig. \ref{fig:Relerror}. The piecewise behavior arises due to the transition occurring at the mixing layer's height, as explained in \cite{DeWisher2014}. \\
\noindent
The first approach assumes that the wind speeds measured at station height $\unit[h_s]{\, [m]}$ are computed from the flat Monin-Obukhov logarithmic models. Using this assumption we interpolate these wind speeds at $h_s$ on a square grid by means of inverse distance weighting. From these interpolated wind speeds we calculate $u_{*}$, $\theta_{*}$, and $Q_{H}$. Using the relationships given in Sect. \ref{sec:translating} the parameters can be transformed to the corresponding  $u_{0_{*}}$, $\theta_{0_{*}}$, and $Q_{H,0}$ for the WKB$_{\epsilon}$ Prandtl models. 
Finally, we use again a relative error function to solve the following optimization problem: 
\noindent
\begin{align}
 \label{eq:f_opt}
 & f(\widehat{u_{0_{*}}}, \widehat{\theta_{0_{*}}},\widehat{u_{*}}, \widehat{\theta_{*}} )= \\ \nonumber
&\Biggl(\left(  \frac{u_{WKB}(h_s \cdot \cos(\alpha)) - u_{MO}(h_s)}{u_{MO}(h_s)}    \right)^{2}+\left(  \frac{\theta_{WKB}(h_s \cdot \cos(\alpha)) - \theta_{MO}(h_s)}{\theta_{MO}(h_s)}    \right)^{2} \\ \nonumber
&+\left(  \frac{u_{0_{*}} - \widehat{u_{0_{*}}}}{u_{0_{*}}}    \right)^{2} +\left(  \frac{\theta_{0_{*}} - \widehat{\theta_{0_{*}}}}{\theta_{0_{*}}}    \right)^{2} +\left(  \frac{u_{*} - \widehat{u_{*}}}{u_{*}}    \right)^{2} +\left(  \frac{\theta_{*} - \widehat{\theta_{*}}}{\theta_{*}}    \right)^{2} \Biggl)^{1/2} \frac{100}{\sqrt{6}} \\ \nonumber
&\longrightarrow \min\limits_{(\widehat{u_{0_{*}}}, \widehat{\theta_{0_{*}}},\widehat{u_{*}}, \widehat{\theta_{*}}  )}. \nonumber 
\end{align}  
\noindent
In above function $u_{*}, \theta_{*}$ are fixed and $u_{0_{*}}, \theta_{0_{*}}$ are given by means of (\ref{eq:theta}) and (\ref{eq:x1}). 
Table \ref{tab:5} gives the calculated parameters for/from the mentioned optimization routine, calculating both the optimal WKB$_{\epsilon}$ Prandtl model and the Monin-Obukhov model. In this example we set $h_s= \unit[5]{\, m}$ and $\alpha=5^{o}$. The relative errors calculated by means of the optimization routine and by means of (\ref{eq:h0}) are small, which shows that our regularization works. Furthermore, the minimum mean relative error consisting of the mean of the relative errors  between the original $u_{0_{*}}, \theta_{0_{*}}$ and the calculated ones is very small, $0.65\%$.  \\
\noindent 
The original values of $u_{{0}_{*}}$ and $\theta_{{0}_{*}}$ for Fig. \ref{fig:Relerror} are given in Table \ref{tab:3}. Corresponding $u_{*}$ and $\theta_{*}$ can be calculated by means of an approach reverse to the one given in Sect. \ref{sec:translating}. 
\begin{table}[ht]
\begin{center}
\caption{Input and output parameters for Fig. \ref{fig:Relerror} using WKB=FALSE \label{tab:5}}
\begin{tabular}{ll|l}
  Par.&Units&Fig. \ref{fig:Relerror} (i/o)\\
  \hline
  $u_{0_{*}}$&$\unit[]{ms^{-1}}$&$0.24/0.24$\\
 $\theta_{0_{*}}$&$\unit[]{K}$&$0.06/0.06$\\
  $u_{*}$&$\unit[]{ms^{-1}}$&$0.21/0.22$\\
 $\theta_{*}$&$\unit[]{K}$&$0.08/ 0.08$\\
  $\sigma_u^2$&$\unit[]{ms^{-1}}$&$0.28/0.28$\\
 $\sigma_w^2$&$\unit[]{ms^{-1}}$&$0.08/0.08$\\
  $\sigma_{u_0}^2$&$\unit[]{ms^{-1}}$&$-/0.27$\\
 $\sigma_{w_0}^2$&$\unit[]{ms^{-1}}$&$-/0.09$\\
$u_{WKB}(h_s \cdot \cos(\alpha))$&$\unit[]{ms^{-1}}$&$-/2.171$\\
$u_{MO}(h_s)$&$\unit[]{ms^{-1}}$&$-/2.174$\\
$\theta_{WKB}(h_s \cdot \cos(\alpha))$&$\unit[]{K}$&$-/269.65$\\
$\theta_{MO}(h_s)$&$ \unit[]{K}$&$-/273.96$\\
$h_0$&$ \unit[]{m}$&$-/$13.79 \\
 $\text{min. mean rel. error}$& $\unit[]{\%}$&$0.65$\\ 
$\text{min. rel. error at } h_0$& $\unit[]{\%}$&$0.52$
\end{tabular}
\end{center}
\end{table}
\noindent 
\newpage
\noindent 
The following is another approach for combining the two models. To begin, establish the following weights for wind speed and potential temperature:
\begin{equation}
\begin{aligned}
\beta(z_{*})&=\mbox{min}\Biggl\{  \frac{\vert u_{WKB}(z)\vert}{\vert u_{MO}(z_{*})\vert},\frac{\vert u_{MO}(z_{*})\vert}{\vert u_{WKB}(z)\vert}\Biggl\}^2,\\
\gamma(z_{*})&=\mbox{min}\Biggl\{  \frac{\vert \theta_{WKB}(z)\vert}{\vert \theta_{MO}(z_{*})\vert},\frac{\vert \theta_{MO}(z_{*})\vert}{\vert \theta_{WKB}(z)\vert}\Biggl\}^2. \label{eq:beta}
\end{aligned}  
\end{equation}  
Second, arrange these weights in a convex combination to calculate total wind speed and potential temperature:
\begin{equation}
\begin{aligned}
u(z_{*})&=&\beta(z_{*})\ u_{WKB}(z)+(1-\beta(z_{*}))\ u_{MO}(z_{*}),\\
\theta(z_{*})&=&\gamma(z_{*})\ \theta_{WKB}(z)+(1-\gamma(z_{*}))\ \theta_{MO}(z_{*}). \label{eq:combined}
\end{aligned}  
\end{equation}  
\noindent
The partial differential equation below is obtained by combining Eq. (\ref{eq:1}) with Navier-Stokes equations and utilising the weights from Eq. (\ref{eq:beta}).
\begin{equation}
\begin{aligned}
0&=\beta(z_{*}) \left(C_2+ K_{H}Pr \frac{\partial^2 u}{\partial z^2} \right) + (1-\beta(z_{*})) \left(K_{H}Pr \frac{\partial^2 u}{\partial z^{*^{2}}} + C_1 \right) \\
&+ \mathbb{I}_{\beta(z_{*}) \leq 0.5} \   \beta(z_{*}) C_1 + \mathbb{I}_{\beta(z_{*}) > 0.5} \ (1-\beta(z_{*})) C_2,
&\\
0&=\gamma(z_{*}) \left( C_3 +   K_{H} \frac{\partial^2 \Delta \theta}{\partial z^2} \right) + (1-\gamma(z_{*})) \left(K_{H}\frac{\partial^2 \theta}{\partial z^{*^{2}}} \right) \\
&+ \mathbb{I}_{\gamma(z_{*}) \leq 0.5} \  \gamma(z_{*}) C_3 + \mathbb{I}_{\gamma(z_{*}) > 0.5} \ (1-\gamma(z_{*})) C_3 \label{eq:combined_ns}
\end{aligned}  
\end{equation} 
with
\begin{equation}
\begin{aligned}
C_1&= \frac{1}{\rho} \frac{\partial p}{\partial z_{*}} + g,\\
C_2&= g \frac{\Delta \theta}{\theta_0} \sin(\alpha), \\
C_3&=  - \left( \Gamma_0 + \epsilon \frac{\partial \Delta \theta}{\partial z}  \right) u \sin(\alpha), \\
\Delta \theta(z) &= \theta(z_{*}) - \theta_0 - \Gamma_0 (z-z_0), \label{eq:combined_delta_theta}
\end{aligned}  
\end{equation} 
\noindent 
and $\mathbb{I}$ is the indicator function. 
 The boundary conditions for our example are selected as follows: $\theta(z_{*}=h_s)= \theta_{WKB}(h_s\cdot  \cos(\alpha))$, $\theta(z_{*}\rightarrow\infty)=\theta(h_{mix})$, $u(z_{*}=h_s)=u_{WKB}(h_s \cdot \cos(\alpha))=u_{MO}(h_s)$, and $u(z_{*}\rightarrow\infty)=u(h_{mix})$. Hence, both wind speed and potential temperature are assumed to remain constant above $h_{mix}$. The mixing layer height is calculated as recommended in \cite{DeWisher2014}.
Equation (\ref{eq:combined_ns}) reduces to the Prandtl system for $\beta=\gamma=1$ and to the Monin-Obukhov system for $\beta=\gamma=0$. A regularized combination of the Prandtl and Monin-Obukhov system is realized if $0< \beta, \gamma < 1$, see Fig. \ref{fig:stoch_navier}.  
Figure \ref{fig:stoch_navier}a describes the WKB$_{\epsilon}$ Prandtl model from Fig. \ref{fig:ex3}a, b, consistently combined according to (\ref{eq:beta}), (\ref{eq:combined}), (\ref{eq:combined_ns}), and (\ref{eq:combined_delta_theta}) with the standard model for flat terrain from Monin-Obukhov similarity theory. Figure \ref{fig:stoch_navier}b states that the errors for Eq. (\ref{eq:combined_ns}) tend to zero, when inserting (\ref{eq:combined}) into the right-hand side of the differential equation, using the weights from (\ref{eq:beta}).
\newpage
 
\newpage


\section{Conclusions}\label{sec:conclusions}
The purpose of this work was to find wind profiles, including slope flows, that can be used to model air dispersion in an alpine valley. 
Slope flows are a major topic in boundary layer meteorology for a variety of reasons. The modification of \cite{Grisogono2015} was used to address these flows, making the general Prandtl model weakly nonlinear. \\
\noindent 
We demonstrated how to obtain friction velocity, friction temperature, and sensible heat flux from the WKB$_{\epsilon}$ Prandtl model and vice versa. Optimization routines were utilised to estimate these values in various function calls, and we demonstrated the effectiveness of our techniques with multiple examples. When the range of $K_{0}$- and $h$ values was too large, problems with applying our R-functions and parameterization occurred because convergence of our methods based on the L-BFGS-B routine and interval bisection was no longer guaranteed. It's worth noting that we tested another optimization routine called SANN, however it was inefficient when compared to L-BFGS-B. \\
\noindent 
One of the main points in this study was the translation of Monin-Obukhov similarity theory friction parameters and sensible heat flux to those for complex terrain with slopes and vice versa. Geometrical analysis, including coordinate rotation, and statistical arguments were used to accomplish this translation.
\newline 
\noindent 
Finally, we discussed the technique to combine the models for flat and steep terrain. This can be achieved using an optimization routine or by solving a variation of the Navier-Stokes equation with a convex combination for total wind speed and potential temperature. For both approaches, we supplied examples. We advise using the second method because of its superior accuracy if the two are compared.
We tested all algorithms in this article to be valid also for different slope angles $\alpha$ between $0^\circ$ and $20^\circ$. The results obtained are comparable to those presented in this article and can be replicated utilizing the code provided in the ancillary files accessible at arXiv.org. For angles exceeding 20 degrees, there exists a significant correlation between the horizontal and vertical wind directions, attributable to the rotation of the coordinate systems. Nevertheless, it is crucial for the model that these wind directions remain uncorrelated. 
From a broader perspective, the results presented in this article can be applicable for simulating pollutant dispersion across the entire micro-meteorology over intricate terrains.
\newpage

\section*{Appendix}\label{secA1}
\begin{table}[ht]
\begin{center}
\caption{Overview of all appearing variables and their associated SI-units \label{tab:appendix}}
\begin{tabular}{lll}
  Variable&Name&Unit\\
  \hline
  $c_{p}=1006$&specific heat capacity& $\unit[]{\, J \ kg^{-1} \ K^{-1}}$\\
  $C$&Prandtl amplitude& $\unit[]{\, K}$\\
  $g=9.81$&acceleration of gravity& $\unit[]{\, m \ s^{-2}}$\\
  $h$&elevation height& $\unit[]{\, m}$\\
  $h_0$&Prandtl layer height& $\unit[]{\, m}$\\
  $h_{mix}$&mixing layer height& $\unit[]{\, m}$\\
  $h_{s}$&station height& $\unit[]{\, m}$\\
  $K_{0}$&constant diffusivity&$\unit[]{\, m^{2} \ s^{-1}}$\\
  $K_{H}$&eddy thermal diffusivity&$\unit[]{\, m^{2} \ s^{-1}}$\\
  $K_{M}$&eddy viscosity&$\unit[]{\, m^{2} \ s^{-1}}$\\
  $L_0$&Obukhov length& $\unit[]{\, m}$\\
  $p$&pressure& $kg \ m^{-1} \ s^{-2}$\\
  $Pr$&Prandtl number&$-$\\
  $Q_{H}$&sensible heat flux&$\unit[]{\, J \ m^{-2} \ s^{-1}}$\\
  $r$&albedo& $-$\\
  $T_{0}$& temperature at surface& $\unit[]{\, K}$\\
  $u,w$& wind speed& $\unit[]{\, m \ s^{-1}}$\\
  $u_{*}$&friction velocity&$\unit[]{\, m \ s^{-1}}$\\
  $z$&orthogonal height above slope&$\unit[]{\, m}$\\
  $z_{j}$&jet height&$\unit[]{\, m}$\\
  $z_{inv}$&temperature inversion height&$\unit[]{\, m}$\\
  $z_{\tiny0}$&roughness length orthogonal to the slope&$\unit[]{\, m}$\\
  $z_{*}$&vertical height above slope &$\unit[]{\, m}$\\
  & & \\
  $\alpha$&slope angle& $\unit[]{\, rad}$\\
  $\Gamma$&background potential  temperature gradient&$\unit[]{\, K \ m^{-1}}$\\
  $\Gamma_0$&background gradient orthogonal to the slope&$\unit[]{\, K \ m^{-1}}$\\
  $\theta$&potential temperature& $\unit[]{\, K}$\\
  $\theta_0$&absolute temperature at roughness length& $\unit[]{\, K}$\\
  $\Delta\theta$&potential temperature anomaly& $\unit[]{\, K}$\\
  $\theta_{*}$&friction temperature &$\unit[]{\, K}$\\
  $\rho=1.2$&density of air & $\unit[]{\, kg  \  m^{-3}}$\\
  $\sigma_u, \sigma_w$&turbulent velocities &$\unit[]{\, m \ s^{-1}}$\\
  $\Phi$& sun elevation angle& $\unit[]{\, rad}$
\end{tabular}
\end{center}
\end{table}
\noindent

\bibliographystyle{spbasic_updated}     
\bibliography{sample_library}

\section*{Statements and Declarations}
\textbf{Funding:} Gunter Sp\"ock and Iris Rammelm\"uller are supported by the Austrian Science Fund (FWF): DOC78.\newline 
\noindent 
\textbf{Author Contributions:} All authors contributed to the study conception and design. Material preparation, data collection and analysis were performed by Maximilian Arbeiter, Iris Rammelm\"uller and Gunter Sp\"ock. All authors read and approved the final manuscript. \newline 
\noindent 
\textbf{Code and Data Availability:} The code and the datasets generated during and/or analysed during the current study are available as ancillary files at arXiv.org \newline 
\noindent 
\textbf{Conflicts of interest/Competing interests:} There is no conflicts of interest and/or competing interests.    \newline 
\noindent 
\textbf{Ethics approval/declarations:} There are no conflicts regarding ethics concerning humans, animals  or anything else in our article or scientific work.    \newline 
\noindent 
\textbf{Consent for publication:} We as authors declare that our article is allowed to be published. We have no conflict of interest.

\end{document}